\documentclass{emulateapj}

\usepackage{hyperref}
\usepackage{xcolor}
\usepackage{url}
\usepackage{amssymb}
\usepackage{physics}
\hypersetup{
    colorlinks,
    linkcolor={red!50!black},
    citecolor={blue!50!black},
    urlcolor={black} 
}
\usepackage{mathrsfs}
\usepackage{nicefrac}

\newcommand{\Jup}{\mathrm{Jup}}

\shorttitle{Reinvestigations of HD 88133 b and ups And b Multi-Epoch Detections}
\shortauthors{Buzard et al.}

\begin{document}

\title{Reinvestigation of the Multi-Epoch Direct Detections of HD 88133 b and Upsilon Andromedae b }

\author{Cam Buzard\altaffilmark{1}, Danielle Piskorz\altaffilmark{2}, Alexandra C. Lockwood\altaffilmark{3}, Geoffrey Blake\altaffilmark{1,2}, Travis S. Barman\altaffilmark{4}, Bj{\"o}rn Benneke\altaffilmark{5}, Chad F. Bender\altaffilmark{6}, John S. Carr\altaffilmark{7}}

\altaffiltext{1}{Division of Chemistry and Chemical Engineering, California Institute of Technology, Pasadena, CA 91125, USA}
\altaffiltext{2}{Division of Geological and Planetary Sciences, California Institute of Technology, Pasadena, CA 91125, USA}
\altaffiltext{3}{Space Telescope Science Institute, Baltimore, MD 21218, USA}
\altaffiltext{4}{Lunar and Planetary Laboratory, University of Arizona, Tuscon, AZ 85721, USA}
\altaffiltext{5}{Institute for Research on Exoplanets, Universit{\'e} de Montr{\'e}al, Montreal, QC, Canada}
\altaffiltext{6}{Department of Astronomy and Steward Observatory, University of Arizona, Tuscon, AZ 85721, USA}
\altaffiltext{7}{Department of Astronomy, University of Maryland, College Park, MD 20742, USA}

\begin{abstract}
We reanalyze the multi-epoch direct detections of HD 88133 b and ups And b that were published in \citealt{Piskorz2016} and \citealt{Piskorz2017}, respectively. Using simulations to attempt to reproduce the detections, we find that with the 6 and 7 $L$ band Keck/NIRSPEC epochs analyzed in the original works, the planets would not have been detectable unless they had unreasonably large radii. HD88133 and ups And both have fairly large stellar radii, which contributed to the difficulty in detecting the planets. We take this opportunity to consider how these planets may have been detectable with the small number of epochs originally presented by running simulations both with the upgraded NIRSPEC instrument and with near-zero primary velocities, as recommended by \citealt{Buzard2021}. While 7 $L$ band NIRSPEC2.0 epochs with near-zero primary velocities would have allowed a strong ($10.8\sigma$) detection of ups And b, many more than 6 $L$ band epochs would have been required for a strong detection of HD88133b, which could be due in part to both this system's large stellar radius and low stellar temperature. This work stresses the importance of careful analytic procedures and the usefulness of simulations in understanding the expected sensitivity of high-resolution spectroscopic data.
\end{abstract}

\section{Introduction}
Direct detection techniques that use radial velocity signatures from exoplanet orbital motion to detect their atmospheric thermal emission have become popular in the last decade. Two variations of such high-resolution techniques have been developed. Both aim to make a direct planetary detection through a measurement of the planetary velocity semi-amplitude, $K_p$. One variation targets planetary systems during periods when the change in the planetary line-of-sight motion is the greatest, typically near conjunction. By observing the system at periods of maximum planetary line-of-sight acceleration, the data contain planetary signatures that shift across the instrument's resolution elements over the course of the night, and techniques like principal component analysis (PCA) can be used to tease apart the changing planetary signal and the stationary telluric and stellar signals. A one-dimensional cross correlation routine can then be used to measure the planetary velocity. This technique was first introduced by \citealt{Snellen2010} with VLT/CRIRES, and has since been applied to data from a range of instruments including Subaru/HDS \citep[e.g.,][]{Nugroho2017}, TNG/GIANO \citep[e.g.,][]{Brogi2018,Guilluy2019}, and CFHT/SPIRou \citep[e.g.,][]{Pelletier2021}.

The second variation of the high resolution technique, which is the focus of this work, instead limits observations so that the change in the planetary line-of-sight velocity is minimized, and the planetary spectrum does not shift across the detector during the course of an observation. This variation is more technically challenging because there is no longer a velocity variation that can be leveraged to separate the planetary and stellar spectra in a single epoch. After the data is telluric corrected, a two-dimensional cross correlation is relied upon to pull apart the stellar and planetary components. Since the planetary signal is so much fainter than the stellar signal, multiple epochs must be combined before the planetary signal becomes apparent. While technically challenging, this variation is the only currently viable high-resolution method for studying the atmospheres of planets whose semi-major axes preclude both single epoch spectroscopic detection, because they move too slowly, and direct imaging with current adaptive optics capabilities, because they are too close to the star ($\lesssim0.1"$, e.g., \citealt{Snellen2014}). This gap includes planets in K dwarf and solar habitable zones. As the so-called multi-epoch technique is uniquely capable of directly studying the atmospheres of non-transiting planets in these systems, it deserves careful work and attention. 




To date, the multi-epoch technique has mainly been applied to data from Keck/NIRSPEC, which is an echelle spectrograph that offered 4-6 orders in the $K$ and $L$ bands per cross disperser setting and R $\sim 25,000 - 30,000$ before its upgrade in early 2019. The method was first applied to Tau Bo{\"o}tis b and was able to measure its $K_p$ as 111 $\pm$ 5 km/s \citep{Lockwood2014}, which was in good agreement with $K_p$ measurements from other techniques \citep[e.g., 110.0 $\pm$ 3.2 km/s, ][]{Brogi2012}. Subsequently, the non-transiting hot Jupiters HD 88133 b and Upsilon Andromedae b (ups And b) were detected at 40 $\pm$ 15 km/s \citep{Piskorz2016} and 55 $\pm$ 9 km/s \citep{Piskorz2017}, respectively. These planets have yet to be studied via a different technique. \citealt{Piskorz2018} then detected the transiting hot Jupiter KELT-2Ab with a $K_p$ of 148 $\pm$ 7 km/s, which was in good agreement with the transit measurement of $145^{+9}_{-8}$ km/s \citep{Beatty2012}. Finally, \citealt{Buzard2020} measured $K_p$ of the non-transiting hot Jupiter HD 187123 b to be 53 $\pm$ 13 km/s. This detection used simulations to identify sources of non-random noise and elucidate the true planetary detection. HD 187123 b has not to date been detected via another technique.

In this work, we look back on the multi-epoch detections of HD 88133 b \citep{Piskorz2016} and ups And b \citep{Piskorz2017}. \citealt{Piskorz2016} reported the Keplerian orbital velocity of HD 88133 b as $40 \pm 15$ km/s using 6 epochs of NIRSPEC $L$ band data and 3 epochs of $K$ band data. \citealt{Piskorz2017} reported the Keplerian orbital velocity of ups And b as $55 \pm 9$ km/s using 7 epochs of NIRSPEC $L$ band data, 3 epochs of $K_l$ band data covering the left-hand half of the NIRSPEC detector, and 3 epochs of $K_r$ band data covering the right-hand half of the detector. In this work, we will focus on the $L$ band data because the $L$ band data provided the majority of the overall structure in both the HD 88133 b and ups And b detections.

\section{Standard Multi-Epoch Analytic Approach}
To begin, we want to give a brief description of the multi-epoch analytic process. These approaches are explained in more detail in prior publications \citep[e.g.,][]{Lockwood2014,Piskorz2016,Buzard2020}. In brief, epochs of data are obtained from hot Jupiter systems over $\sim$2-3 hour periods during which the planetary signal is not expected to significantly shift compared to the wavelength scale of the detector. The two-dimensional echelle spectra are reduced, wavelength calibrated, telluric corrected, and run through a two-dimensional cross correlation routine with appropriate stellar and planetary spectral models. Because the stellar signal is the major component of the data after telluric correction, the known stellar velocity, given by
\begin{equation}
    v_{pri} = v_{sys} - v_{bary}
\end{equation}
where $v_{sys}$ is the systemic velocity and $v_{bary}$ is the barycentric velocity, can always be correctly measured in each epoch. A cut along the known stellar velocity gives a one-dimensional cross correlation in terms of planetary velocity shift. With a very low contrast relative to the stellar signal, the planetary signal requires the combination of multiple epochs to become clearly detectable.

To be combined, the cross correlations must first undergo two transitions. They must first be converted from functions of secondary velocity, which is dependent on orbital phase, to functions of a parameter independent of orbital phase, namely, the Keplerian orbital velocity. Second, they must be converted to log likelihoods. To convert them from functions of secondary velocity to Keplerian orbital velocity ($K_p$), we apply the equation, 
\begin{equation} \label{calcKp}
    v_{sec}(f) = -K_p(\cos(f+\omega_{st}) + e\cos(\omega_{st})) + v_{pri}
\end{equation}
where $f$ is the planet's true anomaly at the observation time, $\omega_{st}$ is the argument of periastron of the star's orbit measured from the ascending node (with the $Z$-axis pointing away from the observer, see \citealt{Fulton2018}), and $e$ is the eccentricity.

To convert the cross correlations from $v_{sec}$ to $K_p$ space using Equation~\ref{calcKp}, we need stellar radial velocity (RV) parameters ($e$, $\omega_{st}$) and true anomaly ($f$) values at each epoch. Stellar radial velocity parameters can typically be found in the literature \citep[e.g.,][]{Butler2006}. We note that it is important that the stellar orbital parameters ($e$, $\omega_{st}$) as well as those used to calculate $f$ ($t_{peri}$, $P$) are pulled from the same literature source. This will be especially important for near-circular orbits where pericenter is not well defined because though references can set pericenter at vastly different points on the orbit, their other parameters (mainly $t_{peri}$ and $\omega$) would then all be consistent to that chosen point of pericenter. A $t_{peri}$ and $\omega$ from different references could be referring to very different points on the orbit, and so could create a large error in the derived $f$s and secondary velocities.

The true anomalies can be calculated using the following equations, which are described in \citealt{MurrayDermott}. First, the mean anomaly ($M$) is calculated from the observation time ($t_{obs}$), and the stellar radial velocity parameters, time of periastron ($t_{peri}$) and orbital period ($P$). If the orbit under study can be assumed circular, the mean anomalies can be used in place of the true anomalies.
\begin{equation}   \label{Equation:M}
    M = 2\pi \bigg{(}\frac{t_{obs} - t_{peri} \mod P}{P}\bigg{)}
\end{equation}
Then, the eccentric anomaly $E$ can be calculated as follows, where $e$ is the eccentricity. As this equation does not have a closed-form solution for $E$ given $M$, $E$ is calculated numerically. 
\begin{equation}
    M = E - e\sin E
\end{equation}
Finally, the true anomaly $f$ is calculated as
\begin{equation}   \label{Equation:f}
    f = 2\arctan\bigg{(}\sqrt{\frac{1+e}{1-e}}\tan\frac{E}{2}\bigg{)}
\end{equation}

We want to make a few important notes about Equations~\ref{calcKp} and \ref{Equation:M}-\ref{Equation:f}. First, the negative sign at the start of Equation~\ref{calcKp}, which is not present in the corresponding equation in \citealt{Fulton2018}, allows this equation to describe the planetary motion, rather than the stellar motion. At any given time, the planetary and stellar motions should have opposite signs. The addition of the negative sign would be equivalent to replacing $\omega_{st}$ with $\omega_{pl}$ (the argument of periastron of the planet's orbit measured from the ascending node) because $\omega_{pl} = \omega_{st} + \pi$ (for a set direction of the $Z$-axis). Importantly, by defining $K_p$ this way, we specify that it must be a positive value. Second, it is important that the zero-point used to define the anomalies ($t_{peri}$ in Equation~\ref{Equation:M}) is consistent with the offset used in Equation~\ref{calcKp}. In the equations as written, $f$ is measured from pericenter, and adding $\omega_{st}$ in Equation~\ref{calcKp} brings the zero point from pericenter to the star's ascending node. The ascending node (for a circular orbit, when the star is at quadrature and moving away from the observer or when the planet is at quadrature and moving toward the observer) should have the largest negative $v_{sec}$ possible. The negative cosine ensures this is the case. Other works that have assumed a circular orbit \citep[e.g.][]{Piskorz2018,Buzard2020} use a positive sine equation with no added phase offset and with $M$s centered at inferior conjunction. This is also a valid approach because the offset between the $M$s and the sine equation are consistent.

Once the planetary cross correlations from different epochs are on the same $K_p$ axis, they must be converted into log likelihoods to be combined. There are a number of ways to do so \citep{Zucker2003,Brogi2019,Buzard2020}. Here, we use the approach first introduced by \citealt{Zucker2003} and termed the ``Zucker ML" approach by \citealt{Buzard2020} to be consistent with \citealt{Piskorz2016} and \citealt{Piskorz2017}. The Zucker ML method converts cross correlations to log likelihoods and combines them as,
\begin{equation}
    \log(L(s)) = \sqrt{1-\exp \bigg{(}\frac{1}{N_{tot}}\Sigma_iN_i\log[1-R_i^2(s)]\bigg{)}},
\end{equation}
where $s$ is the velocity shift, $R_i$ are the individual cross correlations, $N_i$ is the number of pixels per cross correlation, and $N_{tot}$ is the total number of pixels. 

With this basis, we turn to a reinvestigation of previous multi-epoch works by \citealt{Piskorz2016} and \citealt{Piskorz2017}.

\section{HD 88133 B}

\citealt{Piskorz2016} used 6 Keck/NIRSPEC $L$ band and 3 $K$ band epochs to measure the planetary Keplerian orbital velocity of the non-transiting hot Jupiter HD 88133 b as 40 $\pm$ 15 km/s. To do so, they used orbital parameters from their own fit to stellar RV data. This stellar RV data set consisted of 55 RV points; 17 had been previously analyzed by \citealt{Fischer2005} and the rest were new RV points taken by the California Planet Survey with HIRES at the W. M. Keck Observatory. Fitting these data with a Markov Chain Monte Carlo technique following \citealt{Bryan2016}, they reported the orbital parameters (velocity semi-amplitude $K = 32.9 \pm 1.03$ km/s, period $P = 3.4148674^{+4.57e-05}_{-4.73e-05}$ days, eccentricity $e = 0.05 \pm 0.03$, argument of periastron of the star's orbit $\omega_{st} = 7.22^{+31.39}_{-48.11}$$^{\circ}$, and time of periastron $t_{peri} = 2454641.984^{+0.293}_{-0.451}$). 

\subsection{Correction to \citealt{Piskorz2016} Results }

In \citealt{Piskorz2016} there was a systematic error in the implementation of the time of periastron in Equation~\ref{Equation:M} that resulted in a 38.0\% orbital offset in the mean anomalies. The anomalies used in the paper analysis and the corrected anomalies are listed in Table~\ref{epochpositions}. 

\begin{figure}
    \centering
    \noindent\includegraphics[width=21pc]{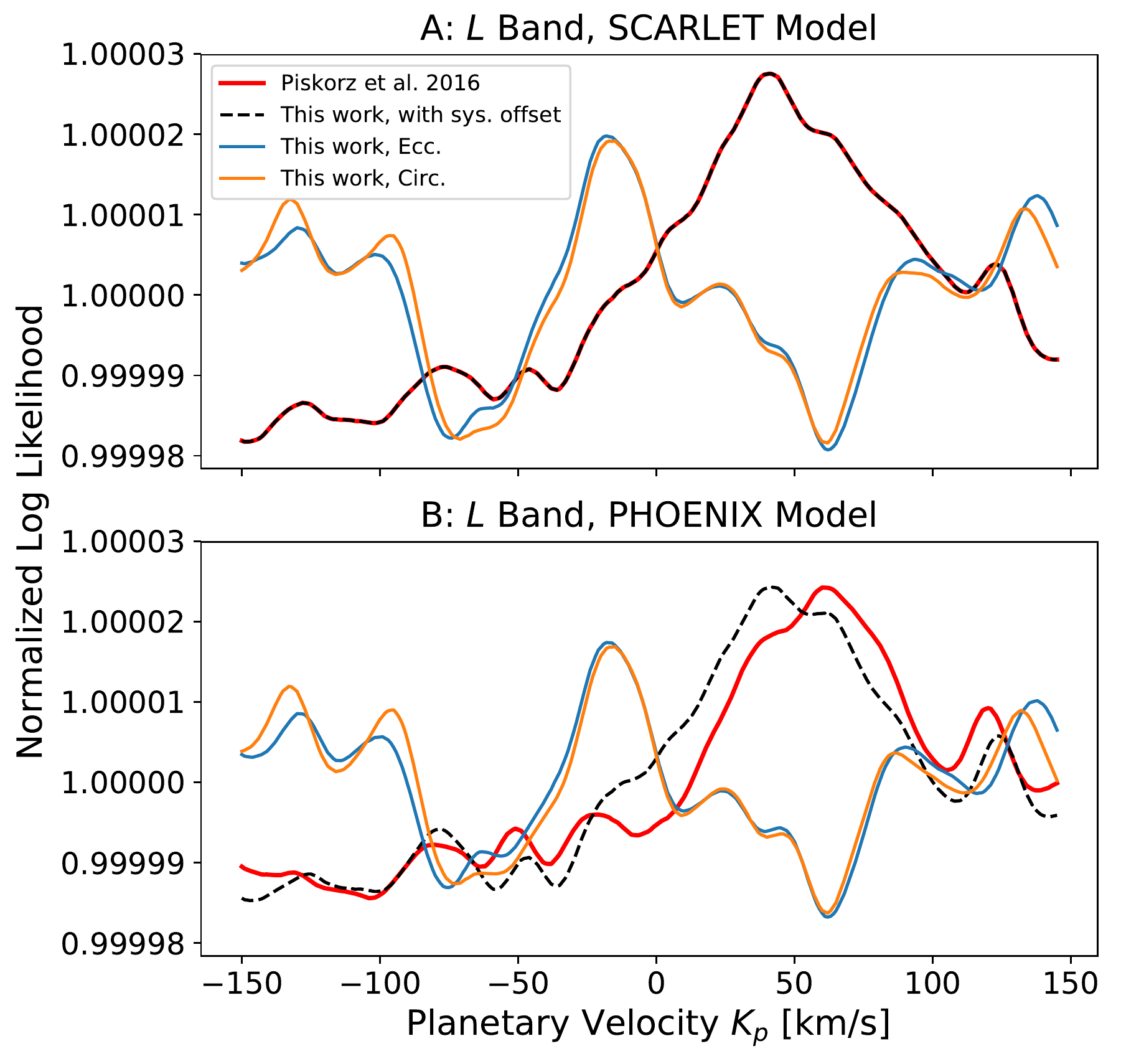}
    \caption{ (A) The normalized log likelihood result for HD 88133 b originally published in \citealt{Piskorz2016} is shown in red. The black dashed curve is able to reproduce the published curve by including a systematic offset in epoch positions. In blue and orange are the corrected log likelihood curves considering an eccentric orbit (blue) and a circular orbit (orange). (B) Same as Panel A, except using a PHOENIX planetary model rather than a SCARLET planetary model. Note here that we were unable to reproduce the published curve. However, when the same orbital parameters are used, we get a much more similar result to that of the SCARLET models than was shown in the \citealt{Piskorz2016}. The two different planetary spectral model frameworks do not create as significant a difference as we thought. }
    \label{hd88133_Lband_logL}
\end{figure}

In Figure~\ref{hd88133_Lband_logL}, we show the corrected log likelihood curves along with the originally published curves, analyzed with both the SCARLET and the PHOENIX planetary models. This correction causes a drastic difference in the resulting log likelihood curve. In each subplot, the red curve was the published log(L) curve and the black dashed curve is the curve reproduced with the systematic offset. Interestingly, this curve exactly reproduces the SCARLET model function, but does not reproduce the PHOENIX model function. In fact, when the data are analyzed with the PHOENIX model, \citealt{Piskorz2016} orbital parameters, and the offset, the resulting function appears much more similar in shape to the corresponding SCARLET result than was originally published in \citealt{Piskorz2016}. This suggests that the two planetary models are much more comparable than was originally thought. 

The blue and orange curves in both subplots show the corrected log likelihoods when the orbit is either treated as eccentric (blue) or assumed to be circular (orange). The similarity between these curves in both subplots shows that for low eccentricity orbits (here $e \sim 0.05$) circular approximations do not greatly affect the shape of the resulting log likelihood surface. 

We also note that when the corrected log likelihood curves show no significant peaks at positive values of $K_p$. Remember that given how we defined the relationship between $v_{sec}$ and $K_p$, only positive values of $K_p$ are physically meaningful. Negative values of $K_p$ would have the stellar and planetary radial velocity curves perfectly in phase rather than out of phase as they should be. This correction therefore implies that we cannot report a measurement of the Keplerian orbital velocity of HD88133b of 40 km/s from the six $L$ band epochs presented in \citealt{Piskorz2016}.

\begin{deluxetable}{lcccc}
\tablewidth{0pt}
\tabletypesize{\scriptsize}
\tablecaption{HD 88133 b Epoch Positions}
\tablehead{ Date  & \multicolumn{2}{c}{\citealt{Piskorz2016}} & \multicolumn{2}{c}{This work}  \\    & $M$\tablenotemark{a} & $f$\tablenotemark{a} & $M$ & $f$ } 
\startdata
 2012 Apr 1 	&		5.11	&	5.01	&	1.22	&	1.31	\\
 2012 Apr 3 	&		2.51	&	2.57	&	4.90	&	4.80	\\
 2013 Mar 10 	&		1.52	&	1.62	&	3.91	&	3.84	\\
 2013 Mar 29 	&		4.95	&	4.85	&	1.06	&	1.15	\\
 2014 May 14 	&		1.03	&	1.12	&	3.42	&	3.40	\\
 2015 Apr 8 	&		3.26	&	3.25	&	5.65	&	5.59	
\enddata
\label{epochpositions}
\tablenotetext{a}{Note, the values here are expressed from 0 to $2\pi$, rather than from 0 to 1 as in \citealt{Piskorz2016}. The $M$ and $f$ values reported in their Table 3 also differ from these values because while they too were affected by the systematic offset, they used \citealt{Butler2006} orbital parameters rather than the newly fit parameters from \citealt{Piskorz2016}.  }
\end{deluxetable}

\subsection{HD88133 Simulations}  \label{Section:HD88133Simulations}

We ran a few sets of simulations to determine what physical or observational factors would allow for the detection of HD88133b. We note that HD88133A has a rather large stellar radius of 2.20 $R_{\sun}$ \citep{Ment2018}, meaning that this system has an even lower planet to star contrast than most hot Jupiters, though we cannot measure it directly because of the unknown planetary radius. Our first simulations ask how large the planetary radius would have to be for the planetary peak to be detectable in the six $L$ band epoch observed (Section~\ref{Section:HD88133PlanaryRadius}). Next, we investigate whether the same epochs taken with the upgraded NIRSPEC2.0 instrument would have been more successful than the NIRSPEC1.0 epochs (Section~\ref{Section:HD88133NIRSPEC2}). Then, we consider the primary velocity. \citealt{Buzard2021} found that epochs with near-zero primary velocities were more useful in damping down structured noise and revealing true planetary signatures than epochs with larger absolute primary velocities. We consider whether the observed epochs would have better revealed the planetary peak if they had smaller absolute primary velocities (Section~\ref{Section:HD88133Vpri}). 

\subsubsection{Generation of Simulated Data}  \label{Section:GenerateSimulations}

For these simulations, we use a stellar model generated from the PHOENIX stellar spectral model grid \citep{Husser2013} interpolated to an effective temperature of 5438 K, a metallicity of 0.330, and a surface gravity of 3.94 \citep{Mortier2013}.  

We use the PHOENIX planetary spectral model from \citealt{Piskorz2016}. This modeled atmosphere does not have an inverted thermal structure in regions close to the molecular photosphere.

We generated the simulated multi-epoch data using the same framework initially presented in \citealt{Buzard2020} and so a full description can be found there. As a quick summary, these simulations combine the stellar and planetary models on the planetary wavelength axis after scaling them by their relative surface areas and shifting them to the desired primary and secondary velocities during each epoch. The secondary velocities are calculated by plugging the $f$ values in Table~\ref{epochpositions}, This work, into Equation~\ref{calcKp} using \citealt{Piskorz2016} orbital parameters and a $K_p$ of 40 km/s. For these simulations, we assume a planetary radius of 1 $R_{\Jup}$ (except in the planetary radius simulations) and a stellar radius of 2.20 $R_{\sun}$ \citep{Ment2018}. After combination, the continuum is removed using a third order polynomial fit from 2.8 to 4.0 $\mu$m in wavenumber space. Then, the simulated data are broadened using the instrumental profiles fit to the data, interpolated onto the data wavelength axes, and the same pixels lost to saturated tellurics in the data are removed. The data, all taken before the NIRSPEC upgrade in early 2019, contain 4 orders per epoch which cover approximately 3.4038–3.4565, 3.2567–3.3069, 3.1216–3.1698, and 2.997–3.044 $\mu$m. Gaussian white noise is added in at the same level as in the data (total S/N per pixel $= 5352$).  

The stellar model used to generate and cross correlate the simulated data differs from the stellar model used to cross correlate the observed data in \citealt{Piskorz2016} in how its continuum is removed. The stellar model used for cross correlation in \citealt{Piskorz2016} was continuum corrected with a second order polynomial fit from 2.0 to 3.5 $\mu$m in wavenumber space, while the model used here is corrected with a third order polynomial fit from 2.8 to 4.0 $\mu$m in wavenumber space. The method of stellar continuum correction actually has a large effect on the shape of the resulting log likelihood curve; when the data are cross correlated with a stellar model corrected by a 3$^{\mathrm{rd}}$ order polynomial fit from $2.8-4.0$ $\mu$m in wavenumber space, the resulting log likelihood curve much more closely resembles the simulated curves (e.g. in Fig.~\ref{planetradiussims}). We use the 3$^{\mathrm{rd}}$ order, $2.8-4.0$ $\mu$m approach in our simulations because this continuum correction method was validated in allowing common structure to be seen in the data and simulated log likelihoods of HD187123b \citep{Buzard2020}. Further, this approach resulted in a flatter looking stellar spectral model, and one that found the known stellar velocities in each epoch of data with higher likelihoods. We do note, however, that the seemingly strong dependence of the final log likelihood shape on the method of stellar model continuum correction is concerning and warrants deeper investigation.

\subsubsection{Planetary Radius Simulations} \label{Section:HD88133PlanaryRadius}

\begin{figure}
    \centering
    \noindent\includegraphics[width=21pc]{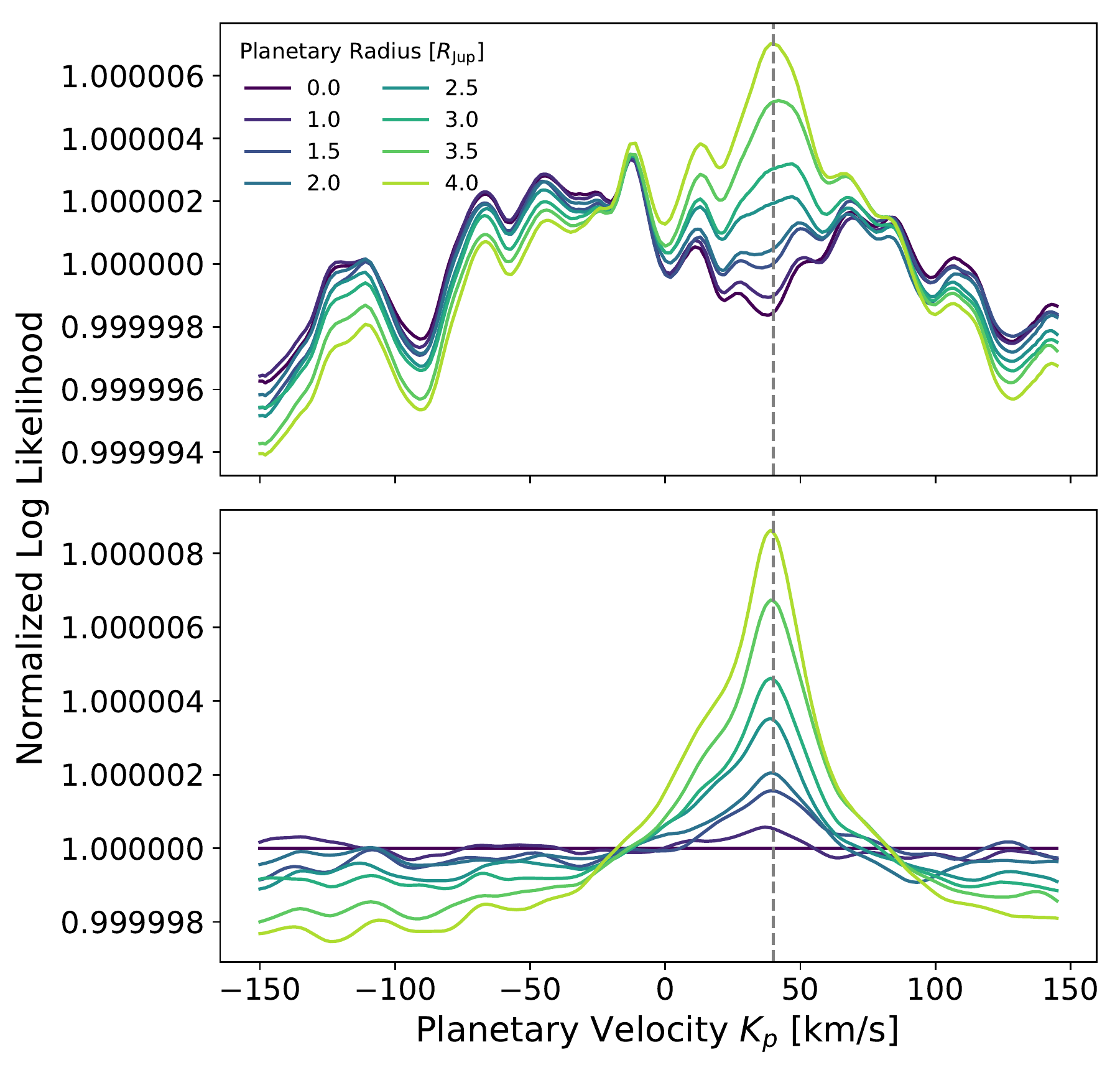}
    \caption{Simulated log likelihood results showing the effects of increasing HD 88133 b planetary radius. These simulations approximate NIRSPEC1.0 $L$ band data taken with the same orbital phases ($f$) and primary velocities in the original data. The bottom panel shows the log likelihood results with the structured noise curve (R$_{pl} = 0$) subtracted off.  }
    \label{planetradiussims}
\end{figure}

Because there seems to be no clear peak at a positive $K_p$ in Figure~\ref{hd88133_Lband_logL}, we use simulations to see how much larger the planetary radius would have to be for its peak to be distinguishable. For these simulations, we set the $K_p$ at 40 km/s and the stellar radius at 2.20 $R_{\sun}$ \citep{Ment2018}, and step the planetary radius up from 1 $R_{\Jup}$ to 4 $R_{\Jup}$ in increments of 0.5 $R_{\Jup}$. Figure~\ref{planetradiussims} shows the results of these simulations in the top panel. In the bottom panel, we show the detections that could be made if all of the structured noise (e.g. the R$_{pl} = 0$ result) could be effectively removed from the results containing a planetary signal.   

In this high S/N per epoch regime, we expect the contribution from structured noise to far outway the contribution from random noise \citep{Buzard2020,Finnerty2021}. The similarity between different radius simulations shows that this is still the case. 

To quantify the strength of these detection peaks, we fit each with a Gaussian model and report the parameters in Table~\ref{Table:HD88133radius}. In the non-corrected versions, the Gaussian model does not fit within one standard deviation of the input $K_p$ until the planetary radius reaches 2.5 R$_{\Jup}$, and the peak does not exceed $3\sigma$ until a radius of 3.5 R$_{\Jup}$. Even at this large radius, the Gaussian model does not clearly distinguish the planetary peak from the structured noise, which can be seen in the Gaussian center offset, large Gaussian width, and relatively low R$^2$ value. Much more promising detections could be made if there were a way to effectively remove the structured noise from the log likelihood results. Even at 1 R$_{\Jup}$, HD 88133 b could have been detected in the data with a significance over $3\sigma$. These simulations take a number of liberties, though, that are not yet applicable to real data. They consider no telluric contamination outside of pixels lost to saturated telluric absorption. They also assume that the stellar and planetary spectra in the data are perfectly matched to the stellar and planetary templates used for cross correlation. Thus, while the corrected simulations shown in the bottom panel of Figure~\ref{planetradiussims} provide an optimistic view of the possible detections with the 6 particular NIRSPEC1.0 $L$ band epochs presented in \citealt{Piskorz2016}, the uncorrected versions give much more realistic estimates.

HD 88133 b has a minimum mass of 0.27 $\pm$ 0.01 $M_{\Jup}$ \citep{Piskorz2016}. With a radius of 3.5 $R_{\Jup}$, it would have a minimum density of 0.01 g/cm$^3$. A growing classifcation of planets with exceptionally large radii for their masses, called ``super-puffs," have low densities of $\lesssim0.3$ g/cm$^3$ \citep[e.g.][]{Cochran2011, JontofHutter2014,Vissapragada2020}. While HD 88133 b ($M_p\sin i \sim 85 M_{\earth}$) is too massive to be classified as a super-puff ($M_p \lesssim 10-15 M_{\earth}$, \citealt{Piro2020}), by comparison of its density, we can conclude that it is highly improbably the planet's radius would be as high as 3.5 $R_{\Jup}$. Indeed, hot Jupiter inflation can approach 2 R$_{\Jup}$ \citep{Thorngren2018}, but has not been observed to exceed it to this extent.

Our simulations therefore confirm that HD 88133 b is not detectable from the six $L$ band epochs of data presented in \citealt{Piskorz2016}. These radius simulations did, however, provide useful information in telling us that the planetary signal would need to be raised by about an order of magnitude (or the structured noise lowered by the same amount), to allow for a confident detection. We now turn to simulations to ask how that order of magnitude may be made up observationally rather than by altering parameters of the physical system like the planetary radius.

\begin{deluxetable}{ccccc} 
\tablewidth{0pt}
\def\arraystretch{1}
\tablecaption{Gaussian Fits to HD88133 Planetary Radius Simulations}
\tablehead{ R$_{pl}$   & $K_p$  & $\Delta K_p$  & Peak Height  & R$^2$  \\ $[$R$_{\Jup}]$  & [km/s] & [km/s] & [$\sigma$] &  }
\startdata
\sidehead{\textbf{Without Star Subtraction}} 
1.0 & 33 & 10 & -0.7 & $2.1\times10^{-3}$ \\
1.5 & -33 & 22 & 1.5 & 0.27 \\
2.0 & -30 & 22 & 1.6 & 0.26 \\
2.5 & 4 & 43 & 2.3 & 0.34 \\
3.0 & 16 & 44 & 2.1 & 0.43 \\
3.5 & 28 & 40 & 3.3 & 0.58 \\
4.0 & 29 & 37 & 5.2 & 0.67 \\
\sidehead{\textbf{With Star Subtraction}} 
1.0 & 35 & 10 & 3.2 & 0.39 \\
1.5 & 40 & 22 & 7.5 & 0.87 \\
2.0 & 36 & 17 & 10.2 & 0.87 \\
2.5 & 34 & 20 & 19.9 & 0.95 \\
3.0 & 36 & 20  & 35.9 & 0.96 \\
3.5 & 37 & 23 & 24.2 & 0.96 \\
4.0 & 35 & 24 & 25.1 & 0.95
\enddata
\label{Table:HD88133radius}
\tablecomments{These simulations were all run with an input $K_p$ of 40 km/s. Prior to fitting, these log likelihood results are subtracted by the mean of their values from -150 to 0 km/s. The Gaussian model is initiated with a 40 km/s center and 10 km/s standard deviation. The Gaussian peak height is reported over $\sigma$, which is measured as the standard deviation of the log likelihood structure more than 3$\Delta K_p$ above or below the best-fit Gaussian center, where $\Delta K_p$ is the standard deviation of the best-fit Gaussian model.  }
\end{deluxetable}

\subsubsection{Upgraded NIRSPEC Simulations} \label{Section:HD88133NIRSPEC2}

The NIRSPEC instrument was upgraded in early 2019, after the \citealt{Piskorz2016} publication. The upgrade would afford 6 usable $L$ band orders per echelle/cross disperser setting as opposed to the 4 from NIRSPEC1.0. It would give twice as many pixels per order (2048 vs. 1024), a nearly doubled spectral resolution ($\sim$41,000 vs. 25,000), and a $\sim$40\% larger wavelength coverage per order \citep{Martin2018}. 

We run NIRSPEC2.0 simulations with the same orbital phases and primary velocities in the original six NIRSPEC1.0 epochs to determine whether the instrument upgrade would have given the planetary signal the boost it needed to be detectable. These simulations are generated as described in Section~\ref{Section:GenerateSimulations}, with the following exceptions. The six orders per epoch cover roughly 2.9331–2.9887, 3.0496–3.1076, 3.1758– 3.2364, 3.3132–3.3765, 3.4631–3.5292, and 3.6349–3.6962 $\mu$m. We broaden the simulated data to an average instrumental resolution of 41,000 and assume a S/N per pixel per epoch of 2860, or a total S/N per pixel of 7000 across the six epochs. The data wavelength axes, locations of saturated telluric pixels, and Gaussian kernals used to broaden the simulated data were taken from the 2019 Apr 3 and 2019 Apr 8 NIRSPEC2.0 data of HD187123 presented in \citealt{Buzard2020}.

\begin{figure}
    \centering
    \noindent\includegraphics[width=21pc]{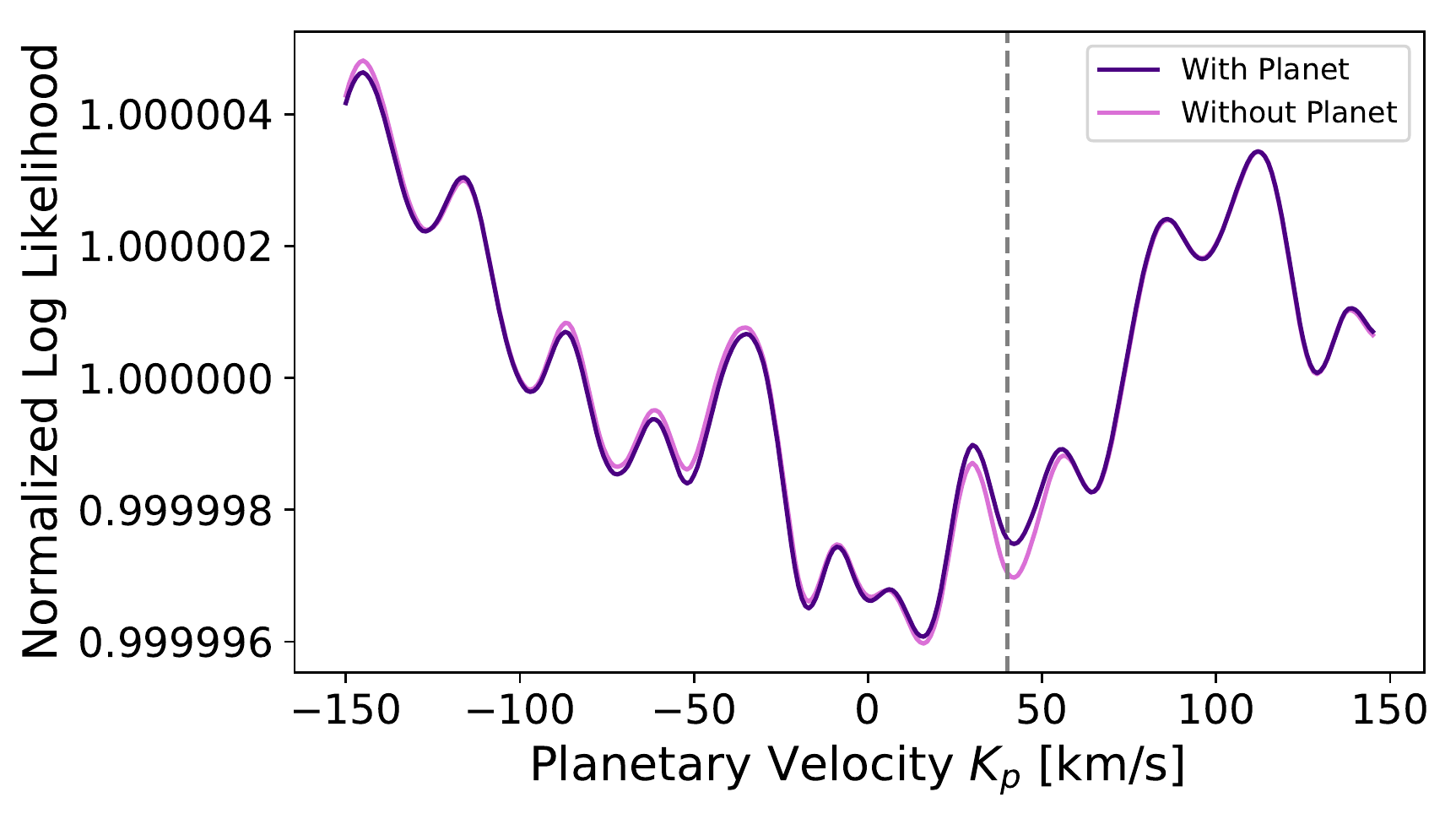}
    \caption{NIRSPEC2.0 simulation of HD 88133 b with same orbital phases ($f$) and primary velocities as in the original data. The data represented by the curve in dark purple has a 1 R$_{\Jup}$ planetary signal and the curve in light purple has no planetary signal.   }
    \label{Figure:NIRSPEC2_HD88133}
\end{figure}

Figure~\ref{Figure:NIRSPEC2_HD88133} shows the results of the simulations which consider imagine the HD88133 $L$ band epochs had been taken with the upgraded NIRSPEC instrument. In light purple is the simulation with no planetary signal added and in darker purple is the simulation with a 1 R$_{\Jup}$ planetary signal. While the likelihood at the input $K_p$ of 40 km/s is increased from the corresponding no planet log likelihood, it does not form a peak and would not constitute a detection. The six $L$ band HD88133 epochs were positioned such that even with the upgraded instrument, they would not have enabled a planetary detection.

\subsubsection{Near Zero Primary Velocity Simulations} \label{Section:HD88133Vpri}

\citealt{Buzard2021} recently showed that, in the small epoch number limit, epochs taken when the primary velocity of a system is near zero are better at reducing structured noise and revealing planetary signatures than epochs taken during periods with larger absolute primary velocities. The majority of the structured noise that arises in the simulations presented here and in \citealt{Buzard2021} results from correlation between the planetary spectral template and the stellar component of the simulated data. We thus suspect that the reduction of structured noise at a primary velocity of zero relates to the portion of the stellar spectrum masked by saturated tellurics when there is a minimal velocity shift between the two spectra. It could be that at this velocity shift, the stellar features that most strongly correlate with the planetary template are masked by saturated tellurics, and without them, the structured noise level decreases substantially. One must be careful in applying this prediction, though, since a primary velocity of zero would bring not just the stellar spectrum, but also the planetary spectrum, closer to the telluric rest frame. While our simulations assume perfect correction of non-saturated tellurics, any residual tellurics that make it through a correction routine could mask planetary features. In a small epoch number limit, an optimal routine might therefore include near-zero primary velocities (to reduce structured noise from star/planet correlation) and quadrature orbital positions (to maximize the planet/telluric velocity separation). With a much larger number of epochs, the structured noise from planet/star correlation may be reduced naturally by the many different combinations of primary and secondary velocities and the usefulness of near-zero primary velocity epochs may be lessened.  

We can ask whether, with just the six epochs on HD88133, near-zero primary velocities might have helped. HD88133 has a primary velocity of zero twice a year: in late February and in mid-August. It would also be accessible from Keck in late February. For the following simulations, we assume epochs had been taken with the same orbital phases ($f$) as the original data, but in late February when $v_{pri} = 0$ km/s. The original data epochs had primary velocities of 17.4, 18.1, 8.1, 16.2, 25.7, and 19.5 km/s. We run these simulations with both the NIRSPEC1.0 and NIRSPEC2.0 configurations.

\begin{figure}
    \centering
    \noindent\includegraphics[width=21pc]{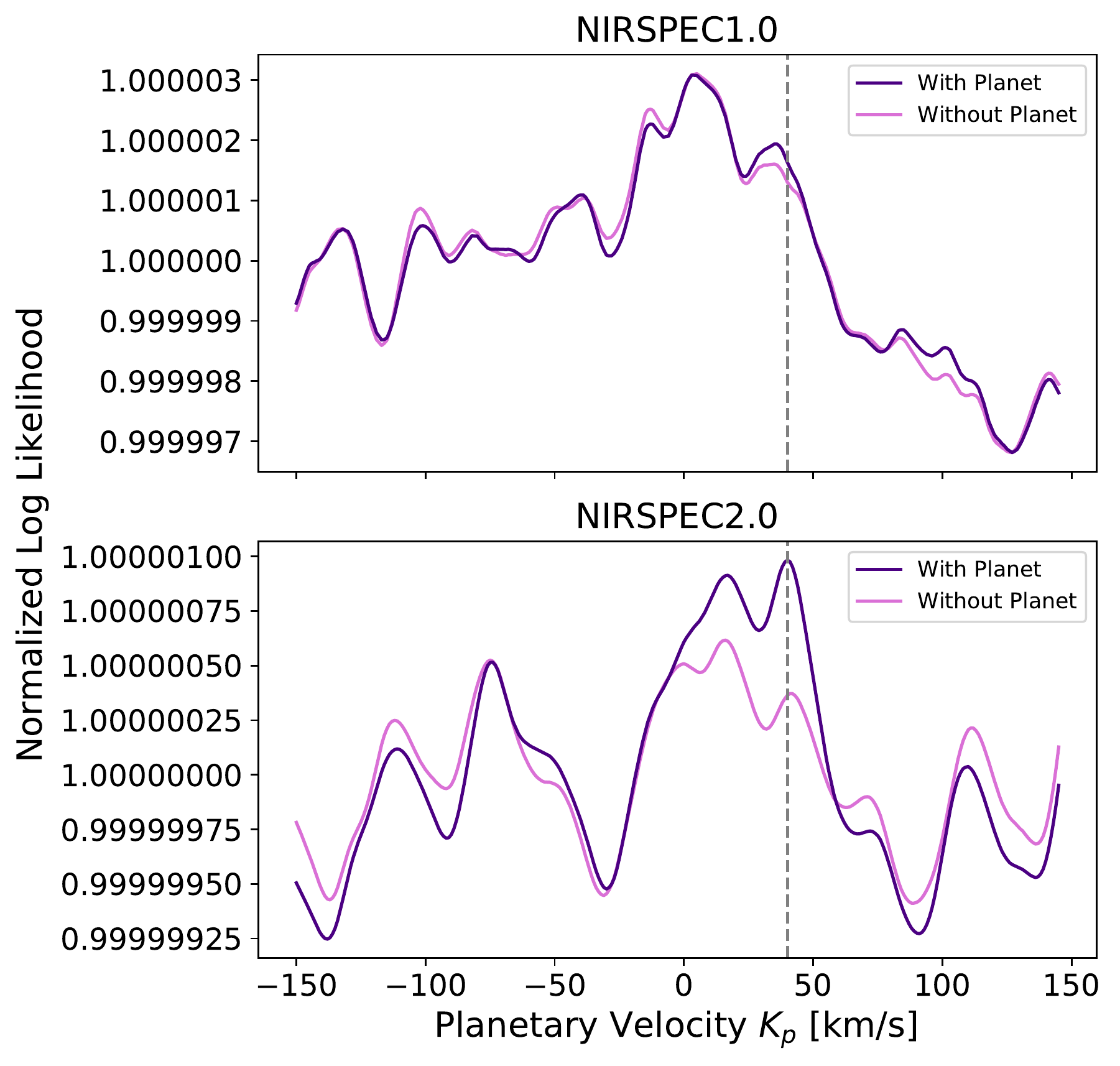}
    \caption{HD 88133 b simulations considering the orbital phases ($f$) from the six original $L$ band epochs, but with 0 km/s primary velocities. The top panel approximated NIRSPEC1.0 data and the bottom panel approximates NIRSPEC2.0 data.   }
    \label{Figure:Nearzeroprimaryvelocity}
\end{figure}

Figure~\ref{Figure:Nearzeroprimaryvelocity} shows the results of the 0 primary velocity simulations with NIRSPEC1.0 results in the top panel and NIRSPEC2.0 results in the bottom panel. In each, we show a pure structured noise simulation (light purple), or simulation with no planetary signal in the simulated data, so that the planetary peak in the simulation with the planetary signal (darker purple) can be distinguished from the structured noise. 

We first consider the NIRSPEC1.0 simulation. While the 1 $R_{\Jup}$ planetary signal definitely shows a larger peak here than when analyzed with the original primary velocities (Fig.~\ref{planetradiussims}), it still does not constitute a very strong detection. We can think of a number of reasons for this. \citealt{Buzard2021} showed that near-zero primary velocity epochs could raise the detection significance on average $\sim2-3\times$ over random primary velocity epochs. From our radius simulations, we estimate an order of magnitude is needed. The gain from near-zero primary velocities then may not be sufficient. HD88133A has an effective temperature of 5438 K \citep{Mortier2013}, putting it on the cooler end of host stars considered by \citealt{Buzard2021}. Cooler host stars showed a stronger preference for near-zero primary velocity epochs, which means in this case, we might expect a bit more than a $3\times$ increase. On the other hand, here, we consider a $K_p$ of 40 km/s, smaller than the 75 km/s $K_p$ used for most of the simulations in \citealt{Buzard2021}. A smaller $K_p$ brings all of the secondary velocities closer in magnitude to the primary velocity; when the primary velocity is 0 km/s, the secondary velocities are closer to 0 km/s, and the planetary spectrum is closer to the telluric rest-frame. That the near-zero primary velocity approach brings the planetary spectrum closer to the telluric frame when combined with a smaller $K_p$ could detract from its advantage over a more random set of primary velocities. Regardless of how these factors work out, Figure~\ref{Figure:Nearzeroprimaryvelocity} confirms that a near-zero primary velocity observing strategy could not have made up for the order of magnitude needed for a strong detection of HD88133b with the orbital phases of the six $L$ band NIRSPEC1.0 epochs obtained and presented in \citealt{Piskorz2016}.

The simulations considering near-zero primary velocity epochs taken with the upgraded NIRSPEC instrument, shown in the bottom panel of Figure~\ref{Figure:Nearzeroprimaryvelocity}, show the most promising chance of detection. There is a peak centered at $K_p = 40$ km/s. A Gaussian fit to the simulated result (dark purple) with no prior information reports a measurement of 22 $\pm$ 20 km/s, with a height of 3.2$\sigma$. If this result were from real data, and we were able to assign the peak at $\sim$17 km/s to noise rather than the planetary signature through either fits with simulations \citep[e.g., ][]{Buzard2020} or because the planet had a radius inflated \citep[e.g.,][]{Charbonneau2000} above the 1 R$_{\Jup}$ assumed here, we could expect a more refined fit.

\section{Upsilon Andromedae b}

\citealt{Piskorz2017} reported the direct detection of upsilon Andromedae b at a $K_p$ of 55 $\pm$ 9 km/s using 7 epochs of Keck/NIRSPEC $L$ band data, 3 epochs of $K$ band data covering the left-hand side of the detector, and 3 $K$ band epochs covering the right-hand side of the detector. For this work, we will again just consider the $L$ band epochs.

\subsection{Correction to \citealt{Piskorz2017} Results } \label{upsandmistake}

\citealt{Piskorz2017} approximated the orbit of ups And b as circular, and so reported mean anomaly $M$ values, rather than true anomaly $f$ values, because they would be the same in the circular limit. We find, however, that there was a systematic error in the calculation of the secondary velocities that stemmed from a mismatch between the zero points used in Equations~\ref{Equation:M} and ~\ref{calcKp}. This resulted in a net error comparable to mean anomalies roughly -3.3\% offset from their true values. Table~\ref{upsandepochpositions} lists the mean anomalies used in \citealt{Piskorz2017} and the corrected anomalies measured from pericenter, calculated using orbital parameters from \citealt{Wright2009}. 

\begin{figure}
    \centering
    \noindent\includegraphics[width=21pc]{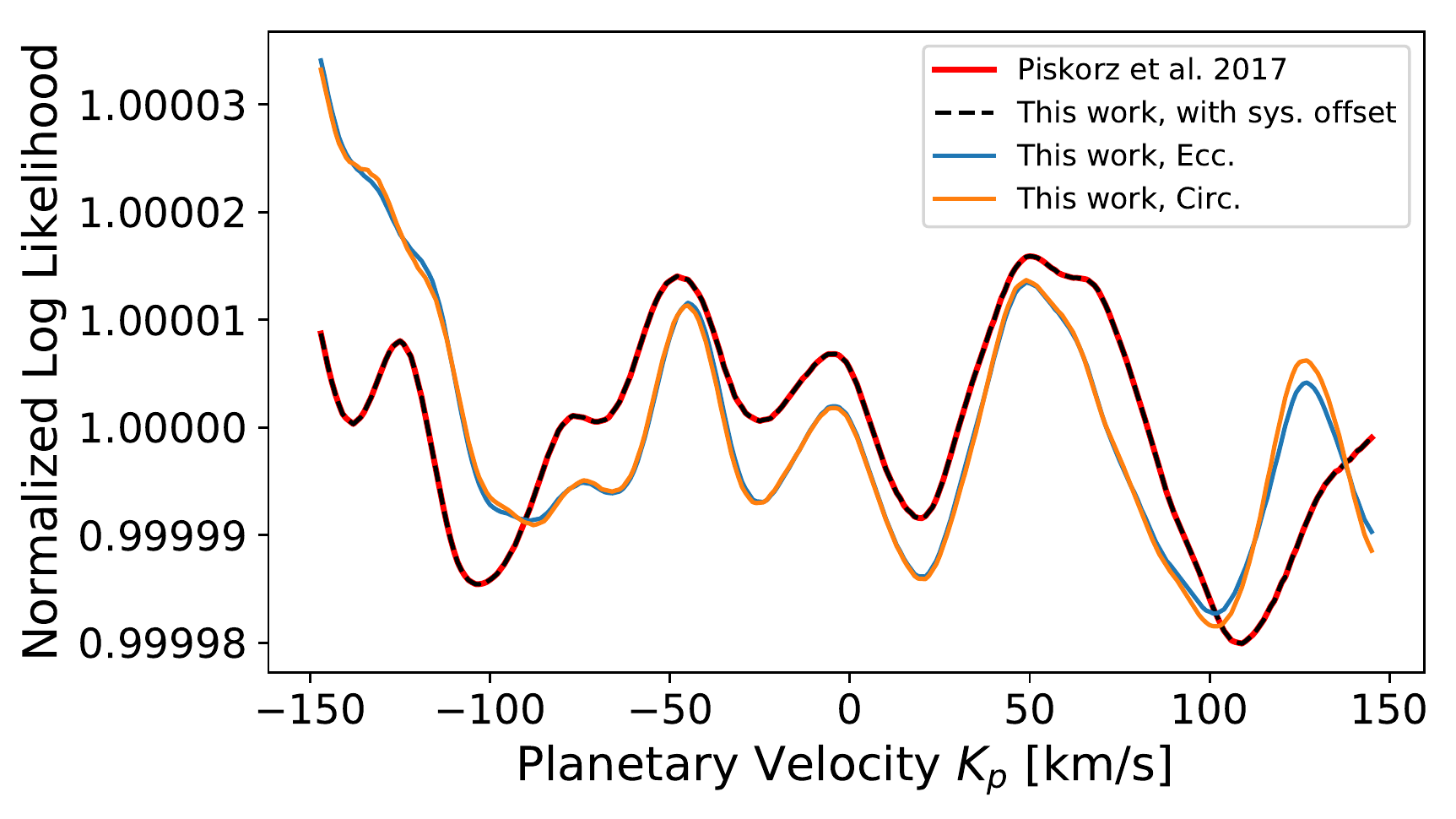}
    \caption{ The normalized log likelihood result for ups And b originally published in \citealt{Piskorz2017} is shown in red. It is reproduced, in black dashed, by including a systematic offset in the epoch positions. The corrected log likelihood curves are shown in blue (eccentric orbit) and orange (circular orbit).   }
    \label{upsand_Lband_logL}
\end{figure}

Figure~\ref{upsand_Lband_logL} shows how these offsets affect the resulting log likelihood curve from the seven epochs of ups And NIRSPEC $L$ band data. The originally published log likelihood curve is in red and is reproduced in black dashed. The corrected log likelihood curves are shown in blue (eccentric orbit) and orange (circular orbit). As in the case of HD 88133 b, we can see here that for low eccentricity orbits, there is no benefit to considering an eccentric orbit rather than assuming a circular one.

\begin{deluxetable}{lccc}
\tablewidth{0pt}
\tabletypesize{\scriptsize}
\tablecaption{Ups And b Epoch Positions}
\tablehead{Date &  \citealt{Piskorz2017}  &    \multicolumn{2}{c}{This work}  \\  &  $M$   & $M$ &$f$ } 
\startdata
2011 Sep 6 &  1.54 &   1.33  & 1.36 \\
2011 Sep 7 &  2.90 &   2.69 & 2.70  \\
2011 Sep 9 &  5.45 &   5.24 & 5.21 \\
2013 Oct 27 &  3.71 &  3.50 & 3.49 \\
2013 Oct 29 &  0.13 &   6.20 & 6.20 \\
2013 Nov 7 &   6.22 &   6.01 & 6.01 \\
2014 Oct 7 &  1.99 &   1.78 & 1.81 
\enddata
\label{upsandepochpositions}
\end{deluxetable}

\subsection{Ups And Simulations}

We were interested in running similar simulations to those run for HD88133 in Section~\ref{Section:HD88133Simulations} to see whether the peak at $\sim$55 km/s can be substantiated. We are particularly curious about whether we could expect the planetary peak to be as strong as it appears in Figure~\ref{upsand_Lband_logL} since ups And, like HD88133, has a large stellar radius \citep[1.7053529$^{+0.1024430}_{-0.0621246}$ R$_{\sun}$,][]{Gaia}. 

\subsubsection{Generation of Simulated Data}

We generate ups And simulated data as described in Section~\ref{Section:GenerateSimulations}. For these simulations, we use a stellar model generated from the PHOENIX stellar model grid \citep{Husser2013} interpolated to an effective temperature of 6213 K, a metallicity of 0.12, and a surface gravity of 4.25 \citep{Valenti2005}. We assume a stellar radius of 1.7053529 R$_{\sun}$ \citep{Gaia} and a planetary radius of 1.0 R$_{\Jup}$ unless otherwise stated. The simulated data are continuum corrected with a 3$^{\mathrm{rd}}$ order polynomial fit from 2.8 to 4.0 $\mu$m in wavenumber space. 

We rotationally broaden the stellar model with a stellar rotation rate of 9.62 km/s \citep{Valenti2005} and limb darkening coefficient of 0.29 \citep{Claret2000}. The FWHM of the instrumental kernels of NIRSPEC1.0 and NIRSPEC2.0 are about 12 and 7.3 km/s, respectively, so while rotational broadening makes little difference to data from NIRSPEC1.0, it would have a larger effect on data from the upgraded NIRSPEC instrument.  

The stellar spectral model used to analyze the ups And $L$ band data in \citealt{Piskorz2017} was not from the PHOENIX grid. Instead, they used a model similar to that described in \citealt{Lockwood2014}. It was generated from a recent version of the LTE line analysis code MOOG \citep{Sneden1973} and the MARCS grid of stellar atmospheres \citep{Gustafsson2008}. Notably, individual abundances were set by matching observed lines for elements that were well measured by NIRSPEC. While tests run on both Tau Boo and ups And NIRSPEC1.0 $L$ band data show that stellar models generated this way are able to measure the true stellar velocities at each epoch with higher likelihoods than PHOENIX stellar models, because these models are generated without a continuum, they cannot be used to generate simulated data. Further, because they rely on fits to NIRSPEC1.0 observational data, they could not be used to generate simulated data outside of the NIRSPEC1.0 order wavelengths. Therefore, we are limited to the PHOENIX stellar model.

We use a planetary model from the SCARLET planetary spectral modeling framework \citep{Benneke2015} without a thermal inversion. The planetary model used in the original work was also from the SCARLET framework, but did include a thermal inversion. We decide to run simulations with a non-inverted planetary model because most recent hot Jupiter atmospheric studies are finding non-inverted thermal structures \citep[e.g., ][]{Birkby2017, Piskorz2018, Pelletier2021}.

We use orbital positions $f$ from the final column of Table~\ref{upsandepochpositions}, \citealt{Wright2009} orbital parameters, and a $K_p$ of 55 km/s in Equation~\ref{calcKp} to determine the secondary velocities at each epoch. The primary velocities at each epoch are -49.7, -49.4, -48.9, -30.5, -29.6, -25.4, and -39.2 km/s. 

Gaussian noise is added at the level of the data (total S/N per pixel $= 18192$).

\subsubsection{Planetary Radius Simulations}

\begin{figure}
    \centering
    \noindent\includegraphics[width=21pc]{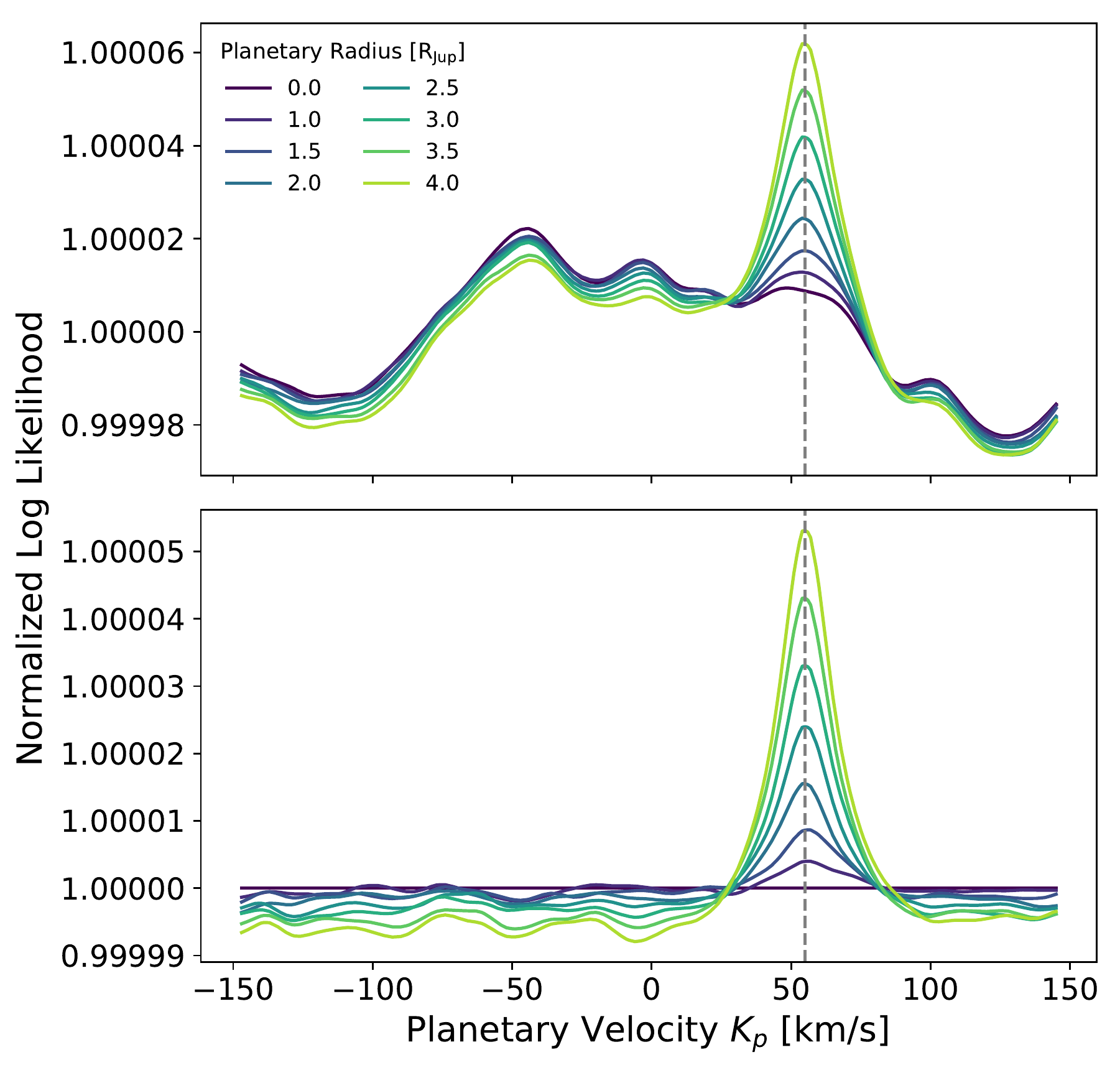}
    \caption{ Ups And simulations showing the effects of increasing planetary radius. The simulations approximate NIRSPEC1.0 $L$ band data with the orbital phases ($f$) and primary velocities from the original 7 epochs of data. The bottom panel shows each log likelihood result with the structured noise curve (R$_{pl} = 0$) subtracted off.  }
    \label{upsand_plradius}
\end{figure}

As for HD88133b, we first run simulations with an increasing planetary radius. Figure~\ref{upsand_plradius} shows the results of these simulations with the planetary radius increasing from 1.0 to 4.0 R$_{\Jup}$ in increments of 0.5 R$_{\Jup}$. Table~\ref{Table:UpsAndradius} report the parameters from Gaussian fits to the log likelihood curves. While Gaussian models can reliably measure a peak centered around the input $K_p$, the R$^2$ values show that a Gaussian model would not be justified until at least a planetary radius of 3.5 to 4.0 R$_{\Jup}$. While transiting hot Jupiters have been observed with radii approaching 2 R$_{\Jup}$ \citep[e.g., KELT-26 b, ][]{RodriguezMartinez2020}, it is improbable that ups And b would have a radius larger than that. These simulations, therefore, do not suggest that, with a reasonable radius, ups And b could be well detected in the 7 original NIRSPEC1.0 $L$ band epochs.

\begin{deluxetable}{ccccc} 
\tablewidth{0pt}
\def\arraystretch{1}
\tablecaption{Gaussian Fits to Ups And Planetary Radius Simulations}
\tablehead{ R$_{pl}$   & $K_p$  & $\Delta K_p$  & Peak Height  & R$^2$  \\ $[$R$_{\Jup}]$  & [km/s] & [km/s] & [$\sigma$] &  }
\startdata
\sidehead{\textbf{Without Star Subtraction}} 
1.0 & 50 & 12 & 0.7 & 0.01 \\
1.5 & 52 & 12 & 1.1 & 0.07 \\
2.0 & 53 & 11 & 1.6 & 0.15 \\
2.5 & 53 & 11 & 2.2 & 0.27 \\
3.0 & 54 & 11 & 2.8 & 0.38 \\
3.5 & 54 & 11 & 3.7  & 0.52 \\
4.0 & 54 & 11 & 4.5 & 0.61 \\
\sidehead{\textbf{With Star Subtraction}} 
1.0 & 58 & 12 & 7.4 & 0.81 \\
1.5 & 56 & 12 & 19.6 & 0.96 \\
2.0 & 55 & 11  & 25.1 & 0.97 \\
2.5 & 55 & 11  & 36.5 & 0.98 \\
3.0 & 55 & 11 & 44.3 & 0.99 \\
3.5 & 55 & 12 & 54.3 & 0.99 \\
4.0 & 55 & 12 & 47.2 & 0.98
\enddata
\label{Table:UpsAndradius}
\tablecomments{These simulations were all run with an input $K_p$ of 55 km/s. Prior to fitting, these log likelihood results are subtracted by the mean of their values from -150 to 0 km/s. The Gaussian model is initiated with a 55 km/s center and 10 km/s standard deviation. The Gaussian peak height is reported over $\sigma$, which is measured as the standard deviation of the log likelihood structure more than 3$\Delta K_p$ above or below the best-fit Gaussian center, where $\Delta K_p$ is the standard deviation of the best-fit Gaussian model.  }
\end{deluxetable}

\subsubsection{Upgraded NIRSPEC Simulations}

The ups And NIRSPEC2.0 simulations are set up the same way as the HD88133 NIRSPEC2.0 simulations with one exception. Because ups And ($K=2.9$) is much brighter than HD88133 ($K=6.2$), we assume a S/N per pixel per epoch of 9000, or a total S/N per pixel of 23800, across the 7 epochs. At the average S/N of 6530 per pixel in the NIRSPEC1.0 data, we were already well into the regime where structured noise far outweighs white noise, so anything more should make little to no difference to the results. 

\begin{figure}
    \centering
    \noindent\includegraphics[width=21pc]{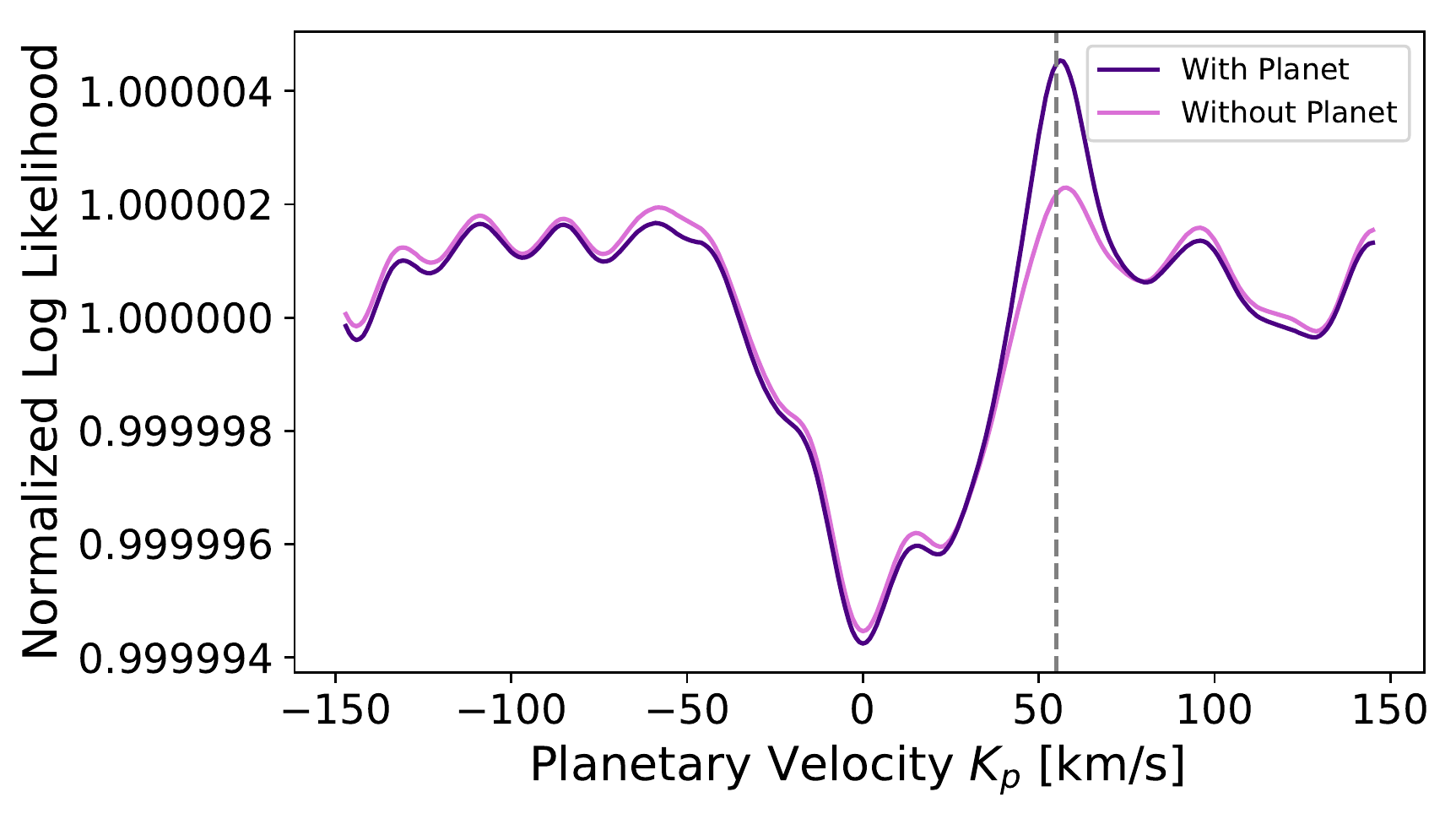}
    \caption{    NIRSPEC2.0 simulation of ups And with the same orbital phases ($f$) and primary velocities as in the original data. }
    \label{Figure:upsand_NIRSPEC2}
\end{figure}

Figure~\ref{Figure:upsand_NIRSPEC2} shows a clear peak at the input $K_p$ of 55 km/s. It does, coincidentally, fall at the same position as a structured noise peak (in light purple), suggesting that its significance could be overestimated. Any other value of $K_p$ would result in a weaker peak that would need to be distinguished, through some mechanism, from the noise peak at $\sim55$ km/s. With the input $K_p$ at 55 km/s, a Gaussian model reports a fit at 57 $\pm$ 7 km/s with a height of 2.1$\sigma$. 

This result is encouraging in that it implies that NIRSPEC2.0 would have allowed a multi-epoch detection of ups And b with the exact seven epochs presented in \citealt{Piskorz2017} even with a planetary radius of 1 R$_{\Jup}$. 

\subsubsection{Near Zero Primary Velocity Simulations}

Because of its relatively large systematic velocity of -28.59 km/s \citep{Nidever2002}, ups And never reaches a primary velocity of 0 km/s. The nearest its primary velocity gets to zero is -2 km/s in late January/early February every year. During this time of year, it would be accessible from Keck during the first few hours of the night, setting, in early February, at around 7 UT. This would optimistically allow for an hour and a half on target after telescope focusing on a good night. Because ups And is a very bright source, enough S/N could be achieved to enter the regime where structured noise, rather than random noise, dominates very quickly. PCA-based telluric correction approaches, like those used in \citealt{Piskorz2016} and \citealt{Piskorz2017}, require enough observation time to witness variation in the telluric spectrum. We run the following simulations assuming that the time before ups And sets would be enough to witness telluric variation sufficient to be picked up by PCA or that the data could be well telluric corrected by another approach. Then, these simulations are run with either the NIRSPEC1.0 or NIRSPEC2.0 set up and with the same orbital phases ($f$) as were in the original data, but with primary velocities at each epoch of -2 km/s.     

\begin{figure}
    \centering
    \noindent\includegraphics[width=21pc]{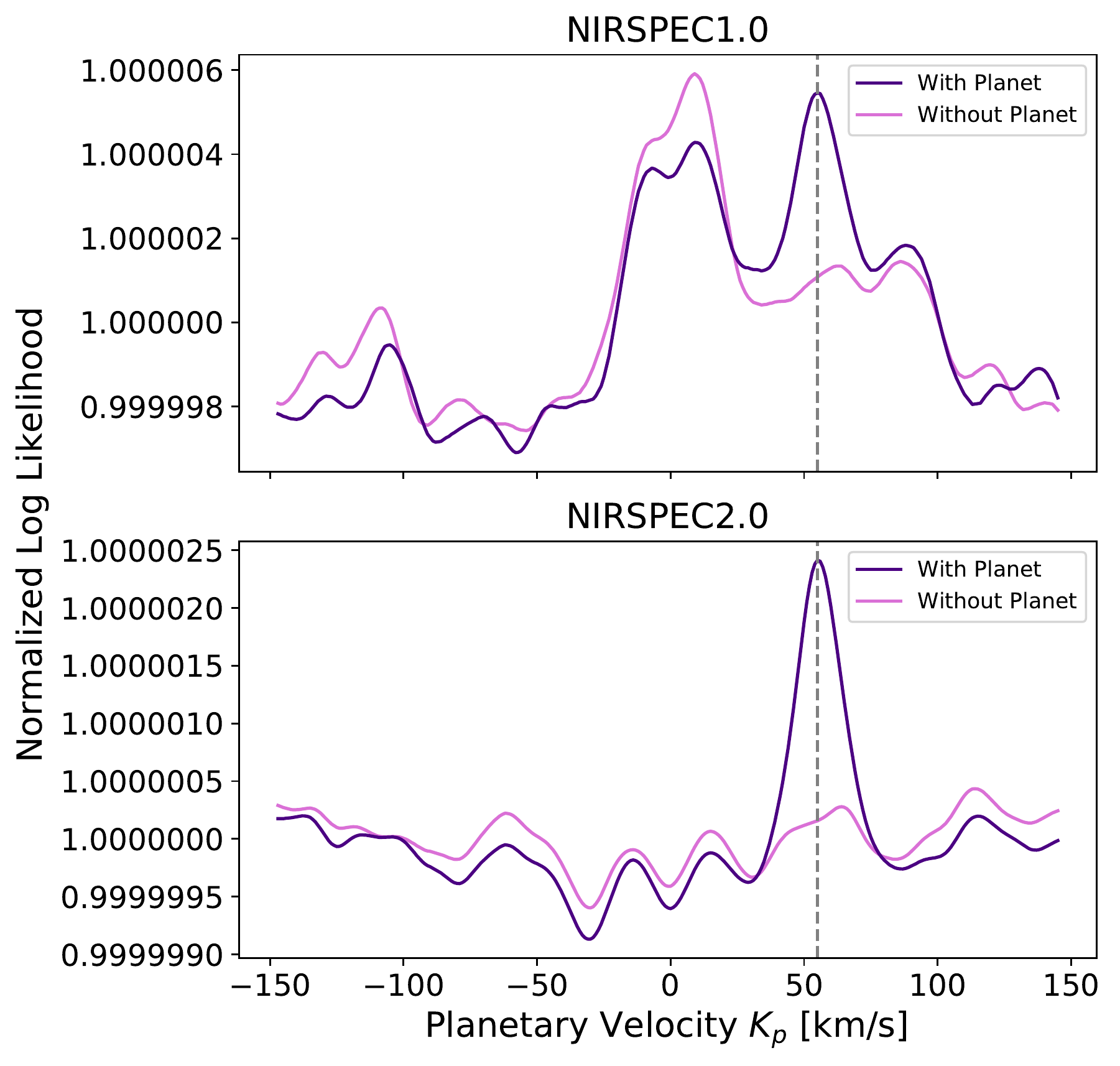}
    \caption{  Ups And simulations with the same orbital phases ($f$) as the original data, but with primary velocities in each epoch equal to $-2$ km/s. The simulations in the top panel approximate NIRSPEC1.0 data and those in the bottom panel approximate NIRSPEC2.0 data.   }
    \label{Figure:upsand_VpriNearZero}
\end{figure}

Figure~\ref{Figure:upsand_VpriNearZero} shows the results of these simulations with the NIRSPEC1.0 results in the top panel and the NIRSPEC2.0 results in the bottom panel. Both configurations show strong features at the input $K_p$ values, with the NIRSPEC2.0 result especially strong and unaltered, in shape, by adjacent structured noise features. A Gaussian model reports a fit to the NIRSPEC2.0 result of $56 \pm 8$ km/s with a height of 10.8$\sigma$. While the 7 $L$ band epochs could have provided a confident planetary detection if taken with NIRSPEC2.0 as is, if they had been taken following the recommendations of \citealt{Buzard2021}, with near-zero primary velocities, they could have presented a very strong detection and a chance for further atmospheric characterization \citep[e.g.,][]{Finnerty2021}.

\section{Discussion}

In this work, we reevaluated the multi-epoch detections of HD 88133 b \citep{Piskorz2016} and ups And b \citep{Piskorz2017}, correcting for errors in the estimated orbital positions at the observation times. Unfortunately, we find that the data is insufficient to report planetary detections or measurements of $K_p$ in either case. Multi-epoch detections with small data sets have always been a risk; stellar radial velocity measurements are now made with tens, if not hundreds, of epochs. 

HD 88133 b and ups And b were two particularly difficult planets to target. Both orbit very large stars, resulting in planet-to-star contrasts lower than those of typical hot Jupiters. At high resolution, the planet/star photospheric contrast provides an upper bound for the spectroscopic information content and thus gives only a first check on how easily detectable planets may be. Predictions of the line-to-continuum, or spectroscopic, contrasts, which can be significantly lower than photospheric contrasts, would provide a more useful guide to direct detection studies; but are highly dependent on the nature of atmospheric chemistry, in particular whether hazes are present, and thermal structure, meaning that model predictions will have large uncertainties. Here we consider photospheric contrasts as a first glimpse into why the spectroscopic detections of HD 88133 b and ups And b may have been so elusive. Measuring photospheric contrast as the mean ratio of planetary flux to stellar flux across the $L$ band, we estimate planet-to-star contrasts for HD 88133 b and ups And b as $2.3\times10^{-4}$ and $2.5\times10^{-4}$, respectively, with an assumed radius of 1 R$_{\Jup}$ for each planet. By comparison, HD 187123 b, studied in \citealt{Buzard2020}, has an expected $L$ band contrast of \textbf{$1.4\times10^{-3}$} and Tau Bo\"otis b, studied in \citealt{Lockwood2014}, has an expected contrast somewhere between 1.1-1.5$\times10^{-3}$, depending on whether or not water is included in its model spectrum (see \citealt{Pelletier2021} for a discussion into water on Tau Boo b). Each of these planets has a contrast nearly an order of magnitude larger than do HD 88133 b and ups And b. Our radius simulations show that in each case, a planetary radius of $\gtrsim$3 R$_{\Jup}$ would have allowed for a strong detection. As planet-to-star contrast increases with R$_{pl}^2$, a radius of 3 R$_{\Jup}$ would increase their contrasts nearly the order of magnitude needed to be comparable to HD 187123 b and Tau Boo b.

We considered whether the upgrade to the NIRSPEC instrument \citep{Martin2018} or the near-zero primary velocity observing strategy presented by \citealt{Buzard2021} would have allowed for enough reduction in structured noise to reveal these low contrast signals. The combination of both would offer a stronger chance of detection in both cases. Near-zero primary velocity epochs obtained with NIRSPEC2.0 would have allowed for $K_p$ measurements of HD 88133 b as $22 \pm 20$ km/s with a height of 3.2$\sigma$ (input $K_p = 40$ km/s) and of ups And b as $56 \pm 8$ km/s with a height of 10.8$\sigma$ (input $K_p = 55$ km/s). Several factors could explain why ups And b could be much more strongly detected under these conditions. It was observed with 7 epochs while HD 88133 b was observed with 6. Additionally, while both stars have large radii, ups And A is not quite as large as HD 88133 A (1.7053529 vs. 2.20 R$_{\sun}$). Perhaps most importantly, ups And A has a higher effective temperature than HD 88133 A (6213 vs. 5438 K). The cooler a stellar effective temperature, the more complex its spectrum will be. Therefore, HD 88133 A would have a more complex spectrum that would allow for more correlation between the stellar component of the data and the planetary spectral model used for cross correlation that gives rise to the structured noise in the final log likelihood results. Additionally, ups And A not only has a less complex spectrum, but also one that is rotationally broadened. The stellar rotational broadening would also work to lessen the degree of correlation between the planetary model and the stellar component of the data. A smaller factor could be the total S/N. Ups And is a much brighter system, and so was simulated with a higher total S/N of 238000 compared to 7000 for HD88133b. Both cases are in a regime where structured noise outweighs white noise, though, so this is not likely a major contributor. Collectively, these factors all work to make ups And b more easily detectable than HD 88133 b. Though, even ups And b was below the detection limit with a small number of NIRSPEC1.0 $L$ band epochs. 

The upgrade to the NIRSPEC instrument will provide a significant advantage to multi-epoch planetary detection, due to its increases in both resolution and spectral grasp \citep{Finnerty2021}. However, in these difficult cases, it might be worthwhile to consider new observational approaches altogether. Since white noise does not appear to be the limiting factor in these multi-epoch studies, we could consider an observational campaign on a smaller telescope (for example, UKIRT) that could dedicate more nights to this work. Another approach could be to consider an instrument like IGRINS or GMTNIRS which could simultaneously afford both a higher spectral resolution than NIRSPEC and a wider wavelength coverage, both of which are beneficial for planet detectability \citep{Finnerty2021}. Many of these instruments are optimized for shorter wavelengths ($<2.5 \mu$m) than we have observed with NIRSPEC. As such, careful work into the optimal observing strategies as well as instrument settings will be crucial in the multi-epoch approach's journey beyond NIRSPEC.

Ultimately, we want to stress the importance of using simulations in multi-epoch work. Simulations are essential for understanding the origin and structure of the expected noise in high resolution data, considering both white noise and any structured noise that may arise in cross correlation space. They can offer realistic estimates for the overall sensitivity of the data beyond expectations from a shot noise limit. As such, simulations can and should be used in many ways, e.g. for planning observations \citep[e.g., ][]{Buzard2021}, for identifying and reducing sources of structured noise \citep{Buzard2020}, and for evaluating approaches of atmospheric characterization \citep[e.g.,][]{Finnerty2021}. 

Despite its challenges, the NIRSPEC multi-epoch approach has been used to characterize planetary atmospheric structure. \citealt{Piskorz2018} combined the multi-epoch detection of KELT-2A b with Spitzer secondary eclipse data. They found that the multi-epoch data provided roughly the same constraints on metallicity and carbon-to-oxygen ratio as the secondary eclipse data. Further, while the secondary eclipse data provided a stronger constraint on $f$, the stellar incident flux which is a rough measure of energy redistribution, the multi-epoch data constrained it to low values, with a 50\% confidence interval at 1.26. As models with $f\gtrsim1.5$ show a temperature inversion, this indicates that using NIRSPEC1.0 multi-epoch data alone, \citealt{Piskorz2018} were able to determine that KELT-2A b has a non-inverted thermal structure in the regions probed by $\sim$3 $\mu$m data. \citealt{Finnerty2021} used NIRSPEC2.0 $L$ band simulations to look more deeply into the atmospheric constraints that could be made with multi-epoch data and found that warm Jupiters' ($T_{\mathrm{eq}} \sim 900$ K) carbon-to-oxygen ratios could be constrained enough to differentiate between substellar, stellar, and superstellar values. While planetary detection using the multi-epoch approach can a challenging pursuit, once the true planetary peak has been identified, the approach holds potential for detailed atmospheric characterization.   

\section{Conclusion}

In this work, we present and correct errors in the multi-epoch detections of HD 88133 b \citep{Piskorz2016} and ups And b \citep{Piskorz2017}. Unfortunately, we find that the original NIRSPEC1.0 $L$ band data presented (6 epochs for HD 88133 b, 7 for ups And b) are insufficient for planetary detections. We run simulations to determine what would have been required for confident detections. Ups And b could have been strongly detected (10.8$\sigma$) if its seven $L$ band epochs had been taken with the upgraded NIRSPEC instrument and following the near-zero primary velocity observing strategy presented by \citealt{Buzard2021}. HD 88133 b would be more difficult to detect, because of its larger stellar radius and lower stellar effective temperature, and would likely have required more, carefully planned, data.

\acknowledgments{ 

This has been a very difficult result to discover and to publish. C.B. wants to acknowledge and thank D.P. for her mentorship. She is an incredible scientist, researcher, and mentor, and his journey through graduate school would not have been possible without her support. As Judith Butler wrote, ``There was and remains warrant ... to distinguish between self-criticism that promises a more democratic and inclusive life for the movement and criticism that seeks to undermine it altogether." We hope this publication falls into the former category and promises a more democratic and inclusive life for the multi-epoch cross-correlation approach moving forward.

We want to acknowledge Katie Kaufman's contribution to the ups And NIRSPEC data reduction along with support from NASA XRP grant NNX16AI14G. We thank our anonymous reviewer for their kind and close reading of our manuscript.  

}

{\footnotesize
\bibliography{sample}}
\bibliographystyle{ApJ}

\end{document}